\documentclass[preprintnumbers,amsmath,amssymb,showpacs]{revtex4}

\usepackage{graphicx}
\usepackage{dcolumn}
\usepackage{bm}
\usepackage{booktabs}
\usepackage{subfigure}
\usepackage{color}
\usepackage{braket}
\usepackage{listings}
\usepackage{soul}
\usepackage{xcolor}
\usepackage{threeparttable}
\usepackage{multirow}
\usepackage{enumerate}
\usepackage{array}
\usepackage{amsthm}
\usepackage{float}
\usepackage{longtable}
\usepackage{epsfig}
\usepackage{mathrsfs}
\usepackage{hyperref}

\soulregister\cite7
\soulregister\ref7
\setstcolor{yellow}

\newcommand{\dd}{\mathop{}\!\mathrm{d}}
\newcommand{\ii}{\mathrm{i}}

\theoremstyle{remark}

\begin{document}

\title{Detecting Nonbilocal Correlations via Parameterized Elegant Joint Measurement}

\author{Ye Tian$^{}$, {Ying-Qiu He$^{\ast}$}, {Zan-Jia Li$^{}$}, {Yu-Yan Zhang$^{}$}, and Dong Ding$^{\dag}$
}
\email{heyq@ncist.edu.cn, dingdong@ncist.edu.cn}

\affiliation{
School of Science, University of Emergency Management, Beijing 101601, China}
\date{\today}

\begin{abstract}
Quantum networks constitute a defining resource for quantum information processing; however, their nonclassical behavior is jointly governed by the choice of measurement basis and the independence of the distributed sources. We study a bilocal entanglement-swapping network in which the central node performs a three-parameter elegant joint measurement (EJM). For two independent Bell-state sources, we construct two binary edge observables and then derive a quantum prediction involving three parameters, which surpasses the classical constraints. In particular, the parameter $\theta$ controls both the degree of entanglement of the EJM basis states and the maximum value of the quantum prediction. These results provide an analytic and directly reproducible route for tuning nonbilocal correlations in an experimentally relevant two-source network.
\end{abstract}


\maketitle

\section{Introduction}

Bell's theorem establishes that quantum correlations cannot in general be explained by a local hidden-variable model \cite{Bell1964,CHSH1969}. This insight has become a central tool in quantum information, where nonlocal correlations support device-independent certification, randomness generation, and cryptographic protocols \cite{Ekert1991,Acin2007,Pironio2010,Brunner2014}. The standard Bell scenario, however, assumes a single common source. Modern quantum information processing is naturally networked: several independent sources distribute systems to spatially separated nodes, and intermediate parties may perform joint measurements that correlate systems which have never directly interacted
\cite{Zukowski1993,Briegel1998,Kimble2008}. In such causal structures, classical explanations must respect not only locality, but also the statistical independence of distinct sources \cite{Fritz2012,Henson2014,Allen2017}.

The bilocal network is the minimal setting in which this additional causal constraint becomes visible. Two independent sources connect Alice with Bob and Bob with Charlie, and Bob performs a joint measurement on the two received systems. Branciard \textit{et al.} showed that source independence leads to nonlinear bilocal inequalities and that entanglement swapping can reveal
nonclassical correlations beyond those detected by treating the two sources as a single hidden variable \cite{Branciard2010,Branciard2012}. Subsequent studies developed the bilocal-network framework in several directions: first by extending it from the basic entanglement-swapping scenario to star-shaped and more general acyclic networks \cite{Tavakoli2014,Tavakoli2016,Andreoli2017,Luo2018}; then by constructing polynomial and network-tailored witnesses suited to these nonlinear causal constraints \cite{Chaves2016,Rosset2016}; and finally by introducing systematic approaches for characterizing and bounding the classical or quantum network correlations \cite{Renou2019,PozasKerstjens2019,Wolfe2021,Supic2020}. Experimental tests have also progressed from proof-of-principle demonstrations to implementations with
independent sources and increasingly strict locality constraints
\cite{Saunders2017,Carvacho2017,Sun2019,Gu2023}.

A distinctive feature of network Bell experiments is that entangled
measurements become as important as entangled states. The standard Bell-state measurement implements entanglement swapping, whereas more general joint measurements can generate network correlations with different noise tolerance, symmetry, and certification properties \cite{Bennett1993,Zukowski1993}.
In 2019, Gisin introduced the elegant joint measurement (EJM), a highly symmetric two-qubit entangled measurement whose local marginals form tetrahedral structures on the Bloch sphere \cite{Gisin2019}. The EJM has since been shown to violate bilocal Bell inequalities \cite{Tavakoli2021}, to generate entanglement-swapping correlations \cite{Huang2022}, and to be relevant for
certifying entangled measurements and quantum-processing tasks
\cite{Baumer2021,Ding2024}. These results make the EJM a useful testbed for understanding how the geometry of a joint measurement is converted into observable network nonlocality.

Recent generalizations have enlarged the EJM from a single symmetric basis to parameterized families of iso-entangled joint measurements \cite{Tavakoli2021,DelSanto2024}. In particular, a three-parameter symmetric EJM introduces the variables $z$, $\phi$, and $\theta$, which respectively affect the  entanglement structure of the measurement basis \cite{He2025}.  Related studies have connected such parameterized joint measurements with bilocal witnesses \cite{Li2026}, experimental network distributions \cite{Wang2026}, and multiqubit elegant measurements
\cite{Ding2025,Pauwels2026,He2026}.
Nevertheless, in a bilocal network it is not immediate which of these
parameters controls the strength of nonbilocality and which merely shifts the phase at which a violation is observed. This distinction is important for analytic comparison with the original EJM.

In this work we analyze a fully specified projective strategy for detecting nonbilocal correlations with the three-parameter EJM. Alice and Charlie each perform measurements with two binary projective observables, while Bob implements the parameterized EJM. We calculate the bilocal correlators and derive the exact
maximum $I_{\max}(\theta)=\sqrt{2\sin\theta}$. We further give the phase conditions for attaining the global value $\sqrt2$ and the condition for violating the bilocal bound.

\section{Preliminaries}

\subsection{Classical bilocal network}

\begin{figure*}[htbp]
  \centering
  \includegraphics[width=0.8\textwidth]{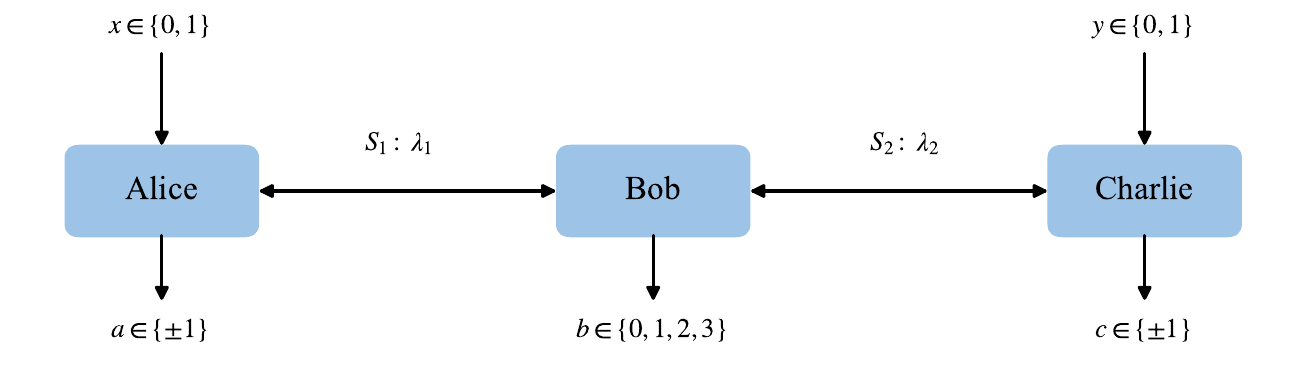}
  \caption{Classical bilocal network considered. The independent sources $S_1$ and $S_2$ distribute hidden variables $\lambda_1$ and $\lambda_2$, respectively. Alice chooses $x\in\{0,1\}$ and outputs $a\in\{\pm1\}$; Charlie chooses $y\in\{0,1\}$ and outputs $c\in\{\pm1\}$; and Bob performs a fixed joint measurement on the two particles, yielding a four-valued outcome labeled $b\in\{0,1,2,3\}$.}
  \label{fig:bilocal-network}
\end{figure*}

The bilocality scenario is the simplest nontrivial network Bell experiment \cite{Branciard2010,Branciard2012}. As
shown in Fig.~\ref{fig:bilocal-network}, two independent sources $S_1$ and $S_2$ connect Alice
with Bob and Bob with Charlie, respectively. In a classical bilocal model, the
source $S_1$ distributes a hidden variable $\lambda_1$ to Alice and Bob, while
the source $S_2$ distributes an independent hidden variable $\lambda_2$ to Bob
and Charlie.
Alice and Charlie choose independent binary inputs $x,y\in\{0,1\}$ and record
binary outputs $a,c\in\{\pm1\}$. Bob performs a fixed two-particle joint measurement, obtaining one of four possible outcomes with $b\in\{0,1,2,3\}$. A joint probability distribution is bilocal if it admits
the decomposition
\begin{equation}
p(a,b,c|x,y)=\int q_1(\lambda_1)q_2(\lambda_2)p(a|x,\lambda_1)p(b|\lambda_1,\lambda_2)p(c|y,\lambda_2)\dd\lambda_1\dd\lambda_2,
\label{eq:bilocal-decomposition}
\end{equation}
where the joint distribution $q(\lambda_1,\lambda_2)=q_1(\lambda_1)q_2(\lambda_2)$ satisfying $\int q_1(\lambda_1) \dd\lambda_1 = \int q_2(\lambda_2) \dd\lambda_2 = 1$.
The product
$q_1(\lambda_1)q_2(\lambda_2)$, rather than an arbitrary joint distribution,
encodes the defining source-independence assumption.

In this case, one usually denotes the dichotomic measurements performed by Alice and Charlie by the observables \(A_x\) and \(C_y\), respectively. Bob's four-valued outcome is commonly characterized by two variables, \(B^0\) and \(B^1\).
Then, the three-particle bilocal distribution satisfies the nonlinear inequality
\begin{equation}
I = \sqrt{|I_1|}+\sqrt{|I_2|}\leq 1,
\label{eq:bilocal-bound}
\end{equation}
where
\begin{equation}
I_1=\frac{1}{4}\sum_{x,y}
\langle A_x C_y B^0\rangle,\quad
I_2=\frac{1}{4}\sum_{x,y}
(-1)^{x+y}\langle A_x C_y B^1\rangle.
\label{eq:bilocal-functionals}
\end{equation}
A value $I>1$ therefore certifies correlations incompatible with the causal model in Eq.~\eqref{eq:bilocal-decomposition}. The conclusion relies on source independence.

\subsection{Parameterized EJM}

Consider the three-parameter two-qubit basis introduced in Ref.~\cite{He2025}. In the computational basis, the four basis states for such three-parameter EJM are represented as:
\begin{equation}
|\Phi_{i}\rangle = \frac{1-\text{i}\sqrt{3z^{2}-1}}{2\sqrt{3}z^{2}}(a_{+}\text{e}^{-\text{i}\phi_{i}}|00\rangle-b_{i,+}^{\theta}|01\rangle
-b_{i,-}^{\theta}|10\rangle+a_{-}\text{e}^{\text{i}\phi_{i}}|11\rangle),\quad i=0,1,2,3,
\end{equation}
where
\begin{equation}
a_{\pm}=\frac{\ii\sqrt{3z^2-1}\pm\sqrt{1-z^2}}{\sqrt2},
\qquad
b^{\theta}_{i,\pm}=\frac{z_i\pm|z|e^{\ii\theta}}{\sqrt2},
\label{eq:coefficients}
\end{equation}
and
\begin{gather}
\phi_0=\phi,\quad \phi_1=\phi+\frac\pi2,\quad \phi_2=\phi-\pi,\quad \phi_3=\phi-\frac\pi2, \quad
z_0=z_2=z, \quad z_1=z_3=-z
\label{eq:phase-signs}
\end{gather}
with $1/{\sqrt3}\leq |z|\leq1, -\pi\leq\phi\leq\pi, 0\leq\theta \leq \pi/2$.
The sign of $z$ changes the phase and outcome convention of the basis. The derivation below uses the positive representative $z\in[1/\sqrt3,1]$; the negative branch gives the same outcomes. We retain $|z|$ in
the final expressions so that the stated parameter domain remains explicit.

\section{Detecting nonbilocal correlations via the three-parameter EJM}

\subsection{Quantum strategy}

We first assume that two independent sources $S_1$ and $S_2$ distribute Bell states
\begin{equation}
\ket{\Phi^+}_{AB_1}=\ket{\Phi^+}_{B_2C}=\frac{\ket{00}+\ket{11}}{\sqrt2}.
\label{eq:source-states}
\end{equation}
The quantum strategy follows
the same source-independence structure as the classical network in
Fig.~\ref{fig:bilocal-network}, but replaces the hidden variables by two
independently distributed Bell states and Bob's classical response by a
projective entangled measurement.

To perform dichotomic measurements, we choose two observables
\begin{equation}
A_0=\sigma_z,\qquad
A_1=\frac{1}{\sqrt{2}}(\sigma_x+\sigma_z)
\label{eq:alice-observables}
\end{equation}
and
\begin{equation}
C_0=\sigma_x,\qquad
C_1=-\frac{1}{\sqrt{2}}(\sigma_x+\sigma_z),
\label{eq:cecil-observables}
\end{equation}
for Alice and Charlie, respectively, where $\sigma_x = |0\rangle \langle 1| + |1\rangle \langle 0|$ and $\sigma_z = |0\rangle \langle 0| - |1\rangle \langle 1|$ are Pauli operators.
For Bob, we define the following two observables constructed from the parameterized EJM projectors:
\begin{equation}\label{eq:bob-eigenvalues}
B^{m}=(-1)^{m}(-|\Phi_{0}\rangle \langle \Phi_{0}|+|\Phi_{1}\rangle \langle \Phi_{1}|+|\Phi_{2}\rangle \langle \Phi_{2}|-|\Phi_{3}\rangle \langle \Phi_{3}|),\quad m=0,1 .
\end{equation}
Therefore, the two terms of the nonlinear inequality \eqref{eq:bilocal-bound} are
\begin{equation}
I_{1}=\frac{1}{4}\langle (A_{0}+A_{1})\otimes(C_{0}+C_{1})B^{0}\rangle,\qquad
I_{2}=\frac{1}{4}\langle (A_{0}-A_{1})\otimes(C_{0}-C_{1})B^{1}\rangle,
\label{eq:I1I2-definition}
\end{equation}
where
\begin{equation}\label{eq:sum-observables}
(A_0+A_1)\otimes(C_0+C_1)=
\frac{1}{2}[(\sqrt{2}-1)\sigma_x\otimes\sigma_x-\sigma_x\otimes\sigma_z
+\sigma_z\otimes\sigma_x-(\sqrt{2}+1)\sigma_z\otimes\sigma_z]
\end{equation}
and
\begin{equation}
(A_0-A_1)\otimes(C_0-C_1)=
\frac{1}{2}[-(\sqrt{2}+1)\sigma_x\otimes\sigma_x-\sigma_x\otimes\sigma_z
+\sigma_z\otimes\sigma_x+(\sqrt{2}-1)\sigma_z\otimes\sigma_z].
\label{eq:difference-observables}
\end{equation}

\subsection{Quantum prediction}

According to the principle of entanglement swapping, with the parameterized EJM taken into account, the four-particle composite system can be written as
\begin{equation}
|\Phi^+\rangle_{AB_1}\otimes|\Phi^+\rangle_{B_2C}
=
\frac12\sum_{i=0}^3
|\widetilde{\Phi}_i\rangle_{AC}\otimes|\Phi_i\rangle_{B_1B_2},
\label{eq:swapping-decomposition}
\end{equation}
where
\begin{equation}
|\widetilde{\Phi}_i\rangle_{AC}
=\frac{1+\text{i}\sqrt{3z^{2}-1}}{2\sqrt{3}z^{2}}(a_{+}^{\ast}\text{e}^{\text{i}\phi_{i}}|00\rangle-(b_{i,+}^{\theta})^{\ast}|01\rangle
-(b_{i,-}^{\theta})^{\ast}|10\rangle+a_{-}^{\ast}\text{e}^{-\text{i}\phi_{i}}|11\rangle)_{AC}.
\label{eq:complex-conjugate-state}
\end{equation}
Therefore, when Bob's joint measurement result is $|\Phi_{i} \rangle$ ($i=0,1,2,3$), we obtain, through further calculation:
\begin{align}
\left\langle\widetilde{\Phi}_{i}\right|
(A_{0}+A_{1})\otimes(C_{0}+C_{1})
\left|\widetilde{\Phi}_{i}\right\rangle
={}&
\frac{1}{4z^{2}}
\Bigg\{
2|z|\sin\theta
\left[
\sqrt{1-z^{2}}\sin\phi_{i}
-\sqrt{3z^{2}-1}\cos\phi_{i}
\right]
\nonumber\\
&\quad
-(\sqrt{2}-1)
\left[
(1-2z^{2})\cos(2\phi_{i})
+\sqrt{1-z^{2}}\sqrt{3z^{2}-1}
\sin(2\phi_{i})
\right]
\Bigg\}
\label{eq:appendix-sum-matrix}
\end{align}
and
\begin{align}
\left\langle\widetilde{\Phi}_{i}\right|
(A_{0}-A_{1})\otimes(C_{0}-C_{1})
\left|\widetilde{\Phi}_{i}\right\rangle
={}&
\frac{1}{4z^{2}}
\Bigg\{
2|z|\sin\theta
\left[
\sqrt{1-z^{2}}\sin\phi_{i}
-\sqrt{3z^{2}-1}\cos\phi_{i}
\right]
\nonumber\\
&\quad
+(\sqrt{2}+1)
\left[
(1-2z^{2})\cos(2\phi_{i})
+\sqrt{1-z^{2}}\sqrt{3z^{2}-1}
\sin(2\phi_{i})
\right]
\Bigg\}.
\label{eq:appendix-difference-matrix}
\end{align}

Based on this, substituting Eqs. \eqref{eq:bob-eigenvalues}, \eqref{eq:sum-observables}, and
\eqref{eq:difference-observables} into Eq. \eqref{eq:I1I2-definition} and taking condition \eqref{eq:phase-signs} into account, we have
\begin{equation}
\begin{split}
I_1=-I_2
=
\frac{\sqrt{2}\sin\theta}{4\,|z|}
\Bigg[
\sqrt{1-z^2}\cos\!\left(\phi+\frac{\pi}{4}\right)
+\sqrt{3z^2-1}\sin\!\left(\phi+\frac{\pi}{4}\right)
\Bigg].
\end{split}
\label{eq:I1I2-final}
\end{equation}
Consequently, the nonlinear bilocal witness becomes
\begin{equation}
\begin{split}
I = \sqrt{|I_1|} + \sqrt{|I_2|}
  = \Bigg[ \frac{\sqrt{2}\sin\theta}{|z|} \Bigg| \sqrt{1-z^2}\cos\left(\phi+\frac{\pi}{4}\right)
  + \sqrt{3z^2-1}\sin\left(\phi+\frac{\pi}{4}\right) \Bigg| \Bigg]^{1/2}.
\end{split}
\label{eq:I-direct}
\end{equation}

\begin{figure*}[htbp]
\centering
\includegraphics[width=0.66\textwidth]{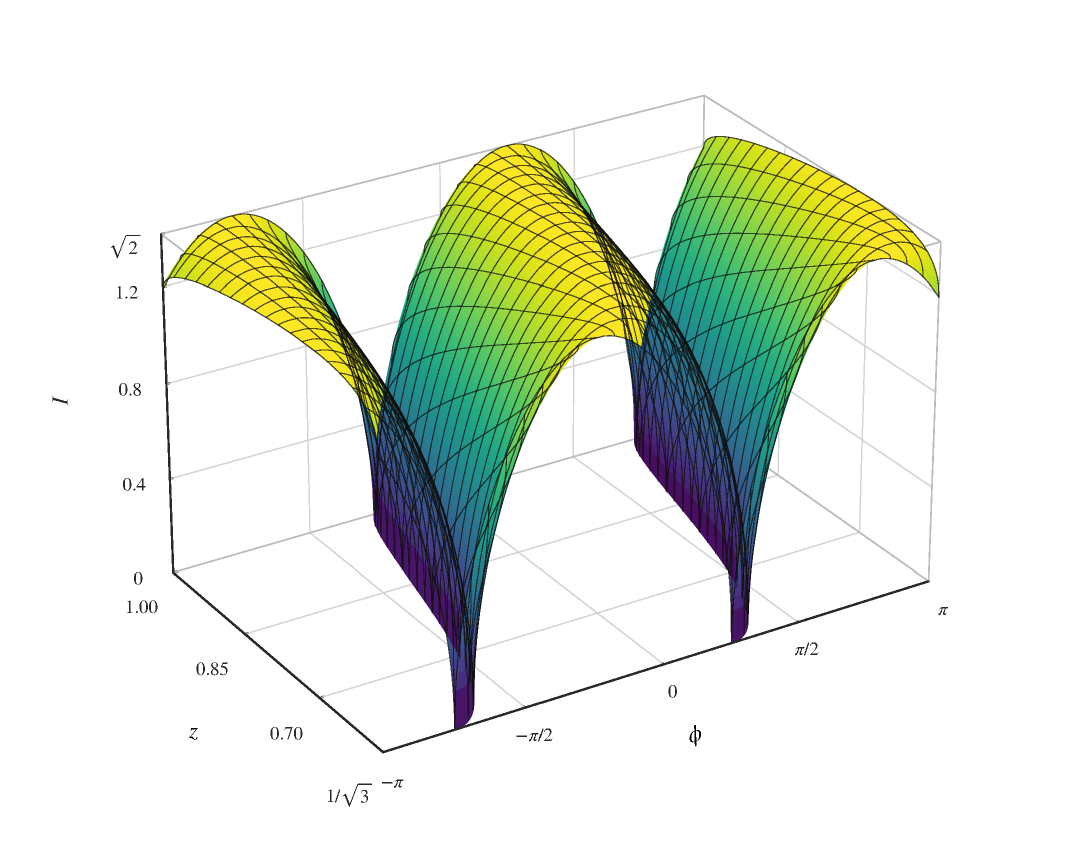}
\caption{Nonlinear bilocal witness $I$ as a function of $z$ and $\phi$ for
$\theta=\pi/2$.}
\label{fig:I-surface}
\end{figure*}

For clarity,  we set the parameter $\theta=\pi/2$ and plot $I$ as a function of $z\in[1/\sqrt3,1]$ and
$\phi\in[-\pi,\pi]$, as shown in Fig. \ref{fig:I-surface}.
Evidently, besides $\theta$, $I$ is related to both variables $z$ and $\phi$ and it ranges over an interval $0\leq I\leq \sqrt{2}$. The two dark valleys correspond to the minimum value $I = 0$, while the yellow ridges reach the maximum value $I = \sqrt2$.

In addition, by simplifying the expression \eqref{eq:I-direct}, we then further derive
\begin{equation}
\begin{split}
I=\sqrt{2\sin\theta\left|
\sin\left[
\phi+\frac{\pi}{4}+
\arctan\left(\frac{\sqrt{1-z^2}}{\sqrt{3z^2-1}}\right)
\right]\right|},
\end{split}
\label{eq:I-compact}
\end{equation}
for $1/\sqrt3 < |z|\le 1$; while it reduces to
$I=\sqrt{2\sin\theta\left|\cos(\phi+\pi/4)\right|}$ for $|z|=1/\sqrt3$.
Since the three parameters $\theta$, $z$ and $\phi$ are independent of each other, as long as $\left|
\sin\left[\phi+\pi/4+\arctan\left({\sqrt{1-z^2}}/{\sqrt{3z^2-1}}\right)\right]\right|=1$ for $1/\sqrt3 < |z|\le 1$ or $|\cos(\phi+\pi/4)|=1$ for $|z|=1/\sqrt3$,  we have
\begin{equation}
I_{\max}(\theta)=\sqrt{2\sin\theta}.
\label{eq:Imax-theta}
\end{equation}
In this case, the parameters $z$ and $\phi$ must satisfy the following conditions
\begin{equation}
\begin{split}
\phi_+&=
\arctan\!\left(\frac{\sqrt{3z^2-1}}{\sqrt{1-z^2}}\right)
-\frac{\pi}{4},
\qquad
\frac{1}{\sqrt{3}}\leq |z|<1,
\end{split}
\label{eq:maximizing_phi-1}
\end{equation}
or
\begin{equation}
\begin{split}
\phi_-&=
\begin{cases}
\arctan\!\left(\dfrac{\sqrt{3z^2-1}}{\sqrt{1-z^2}}\right)
+\dfrac{3\pi}{4},
&
\dfrac{1}{\sqrt{3}}\leq |z|\leq \dfrac{1}{\sqrt{2}} \\[2.0ex]
\arctan\!\left(\dfrac{\sqrt{3z^2-1}}{\sqrt{1-z^2}}\right)
-\dfrac{5\pi}{4},
&
\dfrac{1}{\sqrt{2}}< |z|< 1;
\end{cases}
\end{split}
\label{eq:maximizing_phi-2}
\end{equation}
and $\phi=\pi/4$ or $\phi=-3\pi/4$ for $|z|=1$.

\subsection{Nonbilocal correlations}

\begin{figure*}[htbp]
\centering
\includegraphics[width=0.98\textwidth]{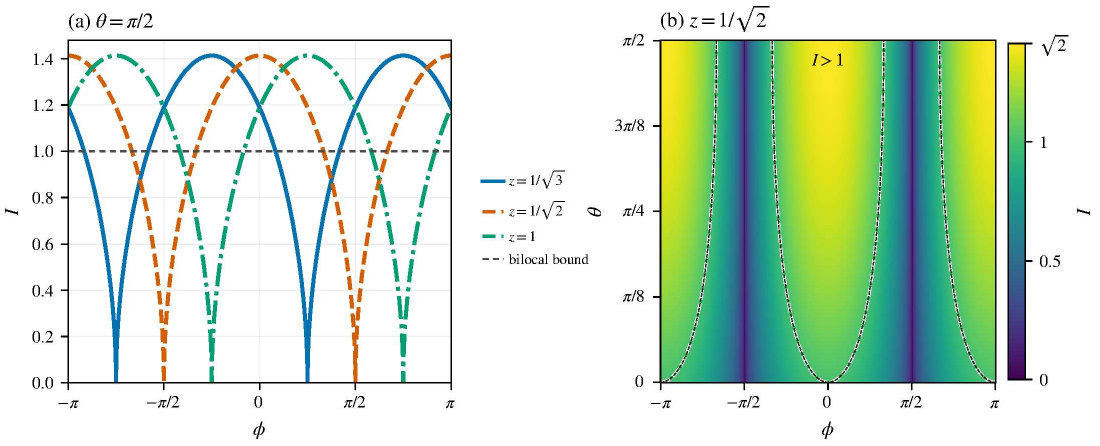}
\caption{Two intuitive illustrations of the nonbilocal correlations. (a) the curves of $I$ as a function of $\phi$ for $z=1/\sqrt{3}$, $z=1/\sqrt{2}$ and $z=1$, respectively. Each curve reaches $\sqrt2$, while $z$ shifts the phase of the extrema. The horizontal dashed line is the bilocal bound $I=1$.
(b) A heatmap of $I$ in the $(\phi, \theta)$ plane. The dashed contour line marks $I=1$, and the quantum violation occurs in the region on its high-value side.}
\label{fig:phase-violation}
\end{figure*}

According to Eq. \eqref{eq:bilocal-bound}, the nonbilocal correlations of the current network can be revealed whenever the value of the right-hand side of Eq. \eqref{eq:I-direct} exceeds 1.
For clarity, we here give two examples. i) We set $\theta = \pi/2$, in which case quantum violation appears as long as $|\sqrt{1-z^2}\cos\left(\phi+\pi/4\right)+\sqrt{3z^2-1}\sin\left(\phi+\pi/4\right)|/|z|
> 1/\sqrt{2}$. More specifically, we further plot the curves of $I$ as a function of $\phi$ for $z=1/\sqrt{3}$, $z=1/\sqrt{2}$ and $z=1$, respectively, as shown in Fig. \ref{fig:phase-violation} (a). Both curves reach a maximum of $\sqrt{2}$, corresponding to different phases. ii) We set $z=1/\sqrt{2}$,
and present the heatmap of $I$ in the $(\phi, \theta)$ plane is shown in Fig. \ref{fig:phase-violation} (b), where the dashed contour line indicates the boundary $I = 1$, above which the quantum violation occurs.

\section{Summary}

In summary, we studied the nonbilocal network correlations based on parameterized EJM.
Our results establish that an multi-parameter joint measurement provides a direct and analytically tractable mechanism for characterizing bilocal correlations in a two-source entanglement-swapping network. Firstly, we constructed a fully specified projective strategy in which Alice and Charlie use fixed binary observables, while Bob performs a single four-outcome parameterized EJM. Secondly, direct evaluation of the quantum correlators gives $I_{1}=-I_{2}$, reducing the nonlinear witness to a single parameter-dependent correlation structure. Thirdly, the three parameters $\theta$, $z$, and $\phi$ jointly determine the value of $I$, while the analytical optimization shows that the maximal attainable value is $I_{\max}(\theta)=\sqrt{2\sin\theta}$. The condition $I>1$ specifies the parameter region in which the bilocal inequality is violated, and the global value $\sqrt{2}$ is reached at $\theta=\pi/2$ with the corresponding choices of the parameters $z$ and $\phi$.

Compared with the recent literature \cite{Li2026}, although both the measurement settings at the network's edge nodes and the resulting types of correlation inequalities differ, the present work, together with that reference, points to and confirms the novel results arising from studying quantum networks based on the parameterized EJM, namely that all three parameters contribute to the quantum predicted values. In particular, when the parameters take special values such that the EJM reduces to the Bell-state measurement, our results are consistent with those of the standard network model based on Bell-basis measurements \cite{Branciard2010,Branciard2012}. These results clarify how a single parameterized measurement at the central node is converted into observable nonbilocal correlations together with the remaining edge measurements. They also provide a closed-form reference for comparing parameterized joint measurements and for investigating the experimental realization of bilocal nonlocality with independent quantum sources.

\begin{acknowledgments}
This work was supported by
the Hebei Natural Science Foundation of China under Grant No. A2026508010
and the Hebei Science and Technology Program Foundation of China under Grant No. 246Z0902G.
\end{acknowledgments}

\vspace*{1cm}

\end{document}